\documentclass[conference]{IEEEtran}

\usepackage{url}
\usepackage[cmex10]{amsmath}
\usepackage{multirow}
\usepackage{epsfig}
\usepackage{mathrsfs}
\usepackage{amssymb}
\usepackage{comment}
\usepackage{bm}
\usepackage{array}
\usepackage{amsthm}
\usepackage{blkarray}
\usepackage{enumerate}
\usepackage{chemarrow}
\usepackage[lined,ruled,linesnumbered]{algorithm2e}
\usepackage{cite}
\usepackage{amsmath,amssymb,amsfonts}
\usepackage{algorithmic}
\usepackage{graphicx}
\usepackage{textcomp}
\usepackage{subcaption}
\usepackage{caption}
\usepackage{makecell}
\usepackage[papersize={8.5in,11in},top=0.75in, bottom=1in, left=0.64in, right=0.64in]{geometry}

\newcolumntype{P}[1]{>{\centering\arraybackslash}p{#1}}
\newcolumntype{M}[1]{>{\centering\arraybackslash}m{#1}}

\newtheorem{theorem}{Theorem}

\newtheorem{definition}[theorem]{Definition}

\IEEEoverridecommandlockouts
\def\BibTeX{{\rm B\kern-.05em{\sc i\kern-.025em b}\kern-.08em
    T\kern-.1667em\lower.7ex\hbox{E}\kern-.125emX}}
\begin{document}

\title{DQ Scheduler: Deep Reinforcement Learning Based Controller Synchronization in Distributed SDN\\
\thanks{This research was sponsored by the U.S. Army Research Laboratory and the U.K. Ministry of Defence under Agreement Number W911NF-16-3-0001. The views and conclusions contained in this document are those of the authors and should not be interpreted as representing the official policies, either expressed or implied, of the U.S. Army Research Laboratory, the U.S. Government, the U.K. Ministry of Defence or the U.K. Government. The U.S. and U.K. Governments are authorized to reproduce and distribute reprints for Government purposes notwithstanding any copyright notation hereon.}
}

\author{\IEEEauthorblockN{Ziyao Zhang\IEEEauthorrefmark{1}, Liang 
		Ma\IEEEauthorrefmark{2}, Konstantinos Poularakis\IEEEauthorrefmark{3}, Kin K. Leung\IEEEauthorrefmark{1}, and Lingfei Wu\IEEEauthorrefmark{2}}
	\IEEEauthorblockA{\IEEEauthorrefmark{1}Imperial College London, London, United Kingdom. Email: \{ziyao.zhang15, kin.leung\}@imperial.ac.uk\\
		\IEEEauthorrefmark{2}IBM T. J. Watson Research Center, Yorktown Heights, NY, United States. Email: \{maliang, wuli\}@us.ibm.com\\
		\IEEEauthorrefmark{3}Yale University, New Haven, CT, United States. Email: konstantinos.poularakis@yale.edu
}}

\maketitle

\begin{abstract}
In distributed software-defined networks (SDN), multiple physical SDN controllers, each managing a network \emph{domain}, are implemented to balance centralized control,  scalability and reliability requirements. In such networking paradigm, controllers synchronize with each other to maintain a logically centralized network view. Despite various proposals of distributed SDN controller architectures, most existing works only assume that such logically centralized network view \emph{can} be achieved with some synchronization designs, but the question of \emph{how} exactly controllers should synchronize with each other to maximize the benefits of synchronization under the eventual consistency assumptions is largely overlooked. To this end, we formulate the controller synchronization problem as a \emph{Markov Decision Process (MDP)} and apply reinforcement learning techniques combined with deep neural network to train a \emph{smart} controller synchronization policy, which we call the \emph{Deep-Q (DQ) Scheduler}. Evaluation results show that DQ Scheduler outperforms the anti-entropy algorithm implemented in the ONOS controller by up to $95.2\%$ for inter-domain routing tasks.
\end{abstract}


\section{Introduction}
Software-Defined Networking (SDN) \cite{kreutz2015software}, an emerging  networking architecture, significantly improves the network performance due to its programmable network management, easy reconfiguration, and on-demand resource allocation, which has therefore attracted considerable research interests.
One key attribute that differentiates SDN from classic networks is the separation of the SDN's data and control plane. Specifically, in SDN, all control functionalities are implemented and abstracted in the \emph{SDN controller}, which sits in the control plane, for operational decision making; while the data plane, consisting of \emph{SDN switches}, only passively executes the instructions received from the control plane. Since the logically centralized SDN controller has full knowledge of the network status, it is able to make the global optimal decision. Yet, such centralized control suffers from major scalability and reliability issues. In this regard, distributed SDN is proposed to balance the centralized and distributed control. 

A distributed SDN network is composed of a set of subnetworks, referred to as \emph{domains}, each managed by a physically independent SDN controller. The controllers synchronize with each other to maintain a logically centralized network view. However, since complete synchronization among controllers, i.e., all controllers always maintain the same global view, will incur high costs especially in large networks, practical distributed SDN networks can only afford partial inter-controller synchronizations, which is known as the \emph{eventual consistency model}\cite{panda2013cap}.  \looseness = -1

The eventual consistency model permits temporarily inconsistent network views among physical distributed controllers in the hope that all controllers will \emph{eventually} be mutually updated. In the mean time, higher network availability, i.e., the ability to provide network services, is realized at the cost of temporary inconsistency according to the CAP theorem \cite{panda2013cap}. Despite the fact that existing works recognize the problems caused by inconsistent network views\cite{levin2012logically}, one crucial question that has been largely overlooked is \emph{precisely how} controllers should synchronize with each other, under limited synchronization budget, to minimize the performance degradation caused by such inconsistency.  For example, ONOS\cite{berde2014onos}, which is a state-of-the-art SDN controller, employs the anti-entropy protocol to realize the eventual consistency\cite{muqaddas2016inter}. The gist of the anti-entropy protocol is that controllers use a simple gossip algorithm to \emph{randomly} synchronize with each other. Although this protocol \emph{can} achieve eventual consistency, is it a \emph{wise} and \emph{efficient} way? \looseness = -1

Motivated by this question and inspired by recent success in applying reinforcement learning (RL) techniques to solve complicated problems, we approach this controller synchronization problem by formulating it as a Markov Decision Process (MDP) problem. Then, we design \emph{Deep-Q (DQ) Scheduler}, an RL based algorithm implemented using Deep Neural Networks (DNN), to decide which controllers to synchronize under the given network synchronization status. The goal of the DQ Scheduler with respect to (w.r.t.) the MDP formulation is to maximize the long-term benefits of controller synchronizations. Evaluations show that DQ Scheduler outperforms the aforementioned anti-entropy protocol by up to $95.2\%$ for the inter-domain routing task.

The rest of the paper is organized as follows. Section~\ref{sec:problem_formulation} formulates the problem and states the objective. Section~\ref{sec:dq_scheduler} describes the design details of the DQ Scheduler. Section~\ref{sec:evaluation} presents the evaluation results of the DQ Scheduler. Section~\ref{sec:related_work} discusses related work. Finally, Section~\ref{sec:conclusion} concludes the paper. \looseness = -1


\section{Problem Formulation}
\label{sec:problem_formulation}

We formulate the controller synchronization problem with inter-domain routing as an application of interest. We first describe the generalized routing path construction mechanism under distributed SDN with eventual consistency model (Section~\ref{path_construction}) and then introduce the performance metric (Section~\ref{performance_metric}). Next, we discuss the synchronization of SDN controllers and introduce its formal definition in Section~\ref{controller_sync}. We then state in Section~\ref{sec:objective} the objective of the controller synchronization problem. Finally, the problem is formulated as an MDP in Section~\ref{sec:MDP_formulation}. 

\subsection{Generalized Path Construction Mechanism in SDN}
\label{path_construction}
Under distributed SDN paradigm, inter-domain routing, like any other network task, is carried out by matching the packet's header with entries in switches' flow tables that store the forwarding rules installed by the controllers. Due to the flexibility and programmability of the SDN, there are potentially many ways in which routing can be conducted. In this section, we describe a simple routing path construction mechanism which is generalized based on principles of BGP-like protocol \cite{kotronis2015routing} in the Internet, and routing mechanisms employed by some state-of-the-art controllers such as the  ONOS controller. Note that it is \emph{not} our intention to design any routing mechanisms; we use this simple and representative mechanism for the sake of problem formulation.  Specifically, the path construction mechanism consists of the following steps.

\emph{\textbf{Step 1}}:  The controller of the domain where the source node sits (\emph{source controller} in the sequel) decides the sequence of domains that the packet will traverse between the source and the destination domains (called the \emph{domain-wise path}), according to certain control objectives of the controller;

 \emph{\textbf{Step 2}}: Based on its view of the topologies of the domains on the domain-wise path, the source controller constructs the path from the source node to the destination node that optimizes the control objective;
 
 \emph{\textbf{Step 3}}: The source controller communicates the path construction decision to the involving domains'  controllers and they install the forwarding rule on switches in the form of ingress and egress gateway IP addresses.

An illustrative example is presented in Fig.\ref{fig:upper_topology} to demonstrate how the routing mechanism works. The example shows the selected domain-wise path between the source node $v_{1}$ and the destination node $v_{2}$, in which domains $\mathcal{A}_{1}$--$\mathcal{A}_{4}$ are involved. The topology of these domains are the views of the source controller, which therefore constructs the path (red lines) that minimizes the hop counts (other performance metrics can also be used; see Section~\ref{performance_metric} for details) between $v_{1}$ and $v_{2}$. Then the source controller instructs the controller of $\mathcal{A}_{2}$ and  $\mathcal{A}_{3}$ that the packet whose destination is $v_{2}$ should egress their domains at node $b$ and $c$, respectively. Note that such constructed path may be suboptimal, as the view is incomplete or out of date (see Section~\ref{controller_sync} for further explanations).

\subsection{Performance Metrics}
\label{performance_metric}
To reach an optimized routing decision under distributed SDN, controllers need to take into account traffic status, load balancing, and other policy-related factors. To this end, controllers can proactively assign a weight to each link to indicate the link preference based on the collected network information, i.e., the smaller the link weight, the better for path construction, so that the end-to-end accumulated weight of any path matches its corresponding path construction preference. Moreover, such link weight assignment is generally adjusted dynamically according to the current network condition. Therefore, the goal for constructing an optimized end-to-end path under a given network condition is reduced to finding the end-to-end path with the minimum accumulated weight under the given link weight assignment. We refer to such accumulated path weight as the \emph{path cost}.


To quantify the performance of the constructed routing path in a selected domain-wise path under the given inter-domain synchronization status, we employ the \emph{Average Path Cost (APC)}, measured by the average cost of the constructed paths, as the performance metric. 

\subsection{Synchronization Among SDN Controllers}
\label{controller_sync}
\begin{figure}
	\centering
\begin{subfigure}[b]{0.4\textwidth}
	\centering
	\includegraphics[width=\linewidth,height=0.25\linewidth]{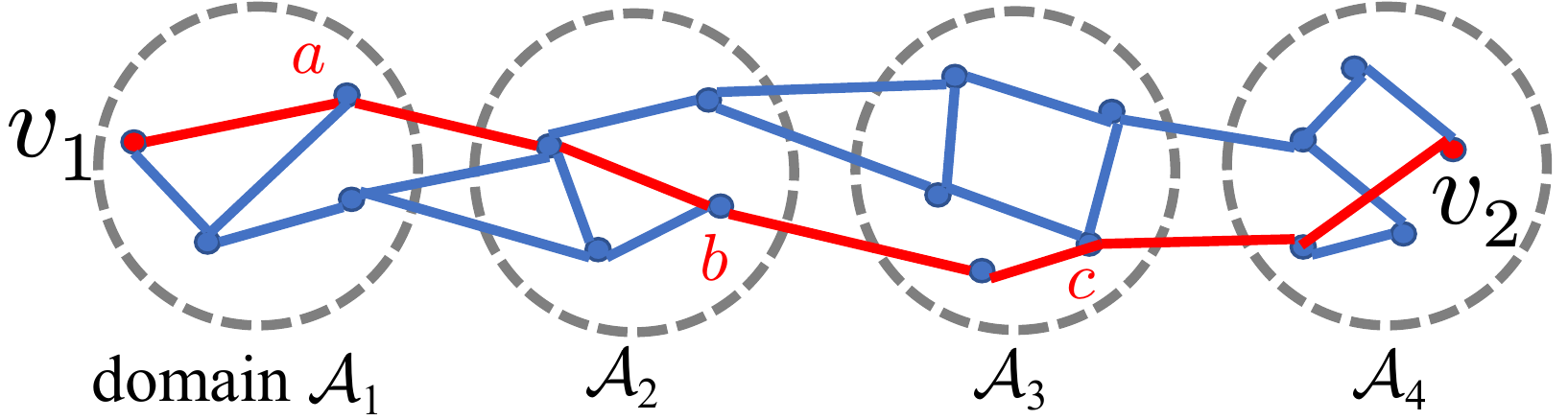}
	\caption{Controller $\mathcal{A}_{1}$'s network view.}
	\label{fig:upper_topology}
\end{subfigure}
\begin{subfigure}[b]{0.4\textwidth}
	\centering
	\includegraphics[width=\linewidth,height=0.25\linewidth]{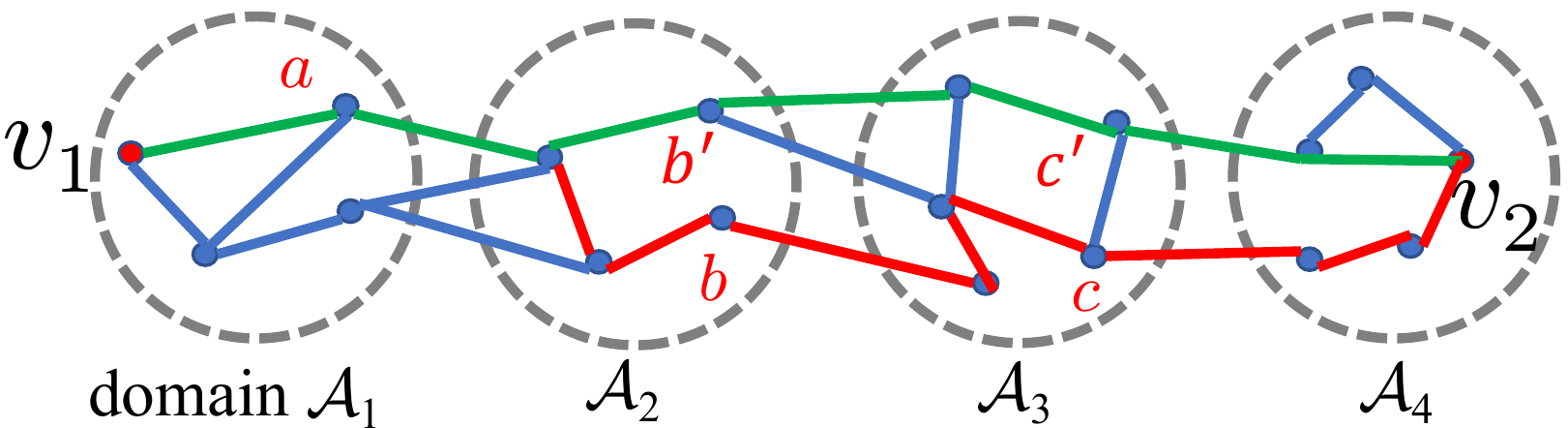}
	\caption{Actual network topology.}
	\label{fig:bottom_topology}
\end{subfigure}
	\label{fig:sync_example}
\caption{A path construction example.}
\end{figure}

Under the eventual consistency model in distributed SDN, the quality of constructed routing paths is directly affected by the controller synchronization levels. We use an example to demonstrate the benefits of controller synchronization for path construction. In Fig.\ref{fig:upper_topology} and Fig.\ref{fig:bottom_topology}, suppose the source node $v_{1}$ in $\mathcal{A}_{1}$ sends packets to the destination node $v_{2}$ in $\mathcal{A}_{4}$. The topology in Fig.\ref{fig:upper_topology} represents $\mathcal{A}_{1}$'s controller's view of the network, which was obtained during synchronizations between $\mathcal{A}_{1}$ and $\mathcal{A}_{2}$--$\mathcal{A}_{4}$ in the past. However, due to the dynamicity of the networks, the actual topology evolves into the one in Fig.\ref{fig:bottom_topology}, which is not promptly synchronized to the source controller. As a result, the source controller, with the outdated view of the network, still uses the old flow table entries which direct packets sent to $v_{2}$ to gateways $a$, $b$, and $c$, respectively. In comparison, the source controller will select a shorter route (green lines) that  involves $b'$ and $c'$ as egress gateways in domains $\mathcal{A}_{2}$, and $\mathcal{A}_{3}$, should it obtain the most up-to-date network topology through synchronizations. This example highlights the important role of \emph{controller synchronization} in dynamic networks, which is formally defined below.

%


\begin{definition}
	\label{def:synch}
	Domain $\mathcal{A}_i$ is synchronized with domain $\mathcal{A}_j$ if and only if the SDN controller in $\mathcal{A}_i$ knows the minimum path cost between any two nodes in $\mathcal{A}_j$.
\end{definition}
Furthermore, we also define the synchronization budget which limits the amount of controller synchronizations. 

\begin{definition}
	\label{def:sync_budget}
	The synchronization budget of an SDN controller is defined as the maximum number of other controllers that it can synchronize with at any time slot.
\end{definition}

\subsection{Objective}
\label{sec:objective}
Two questions motivate our definition of problem objective. First, how does the source controller make synchronization decisions that most efficiently utilize the limited synchronization budget? Second, how to maximize the benefit of synchronization over time? With these questions in mind, we formally state the objective of the controller synchronization problem. \looseness = -1

\emph{\textbf{Objective}}: In dynamic networks whose topologies evolve over time, given the  controller synchronization budget and for a set of source and destination nodes located in different domains that send/receive data packages, how does the source controller synchronize with other controllers on the domain-wise path at each time slot, to maximize the benefit of controller synchronization (reductions in APC for delivering these packets) over a period of time? 


\subsection{MDP Formulation}
\label{sec:MDP_formulation}
We formulate the controller synchronization problem as a Markov Decision Process (MDP), in which  3-tuple $(\mathcal{S},\mathcal{A},R)$ is used to characterize it.  

\begin{itemize}
	\item  $\mathcal{S}$ is the finite state space. In our problem, a state corresponds to the counts of time slots since the source controller was last synchronized with other controllers on the domain-wise path. 
	\item $\mathcal{A}$ is the finite action space. An action w.r.t. a state is defined as the decision to synchronize with the selected domain(s), subject to the given synchronization budget.  
	\item $R$ represents the immediate reward associated with state-action pairs, denoted by $R(s,a)$, where $s \in \mathcal{S}$ and $a \in \mathcal{A}$. $R(s,a)$ is calculated as the average reductions in APC associated with an $(s,a)$ tuple.
\end{itemize}

\begin{figure}
	\smallskip
	\centering
	\includegraphics[height=0.55\linewidth,width=3.4in]{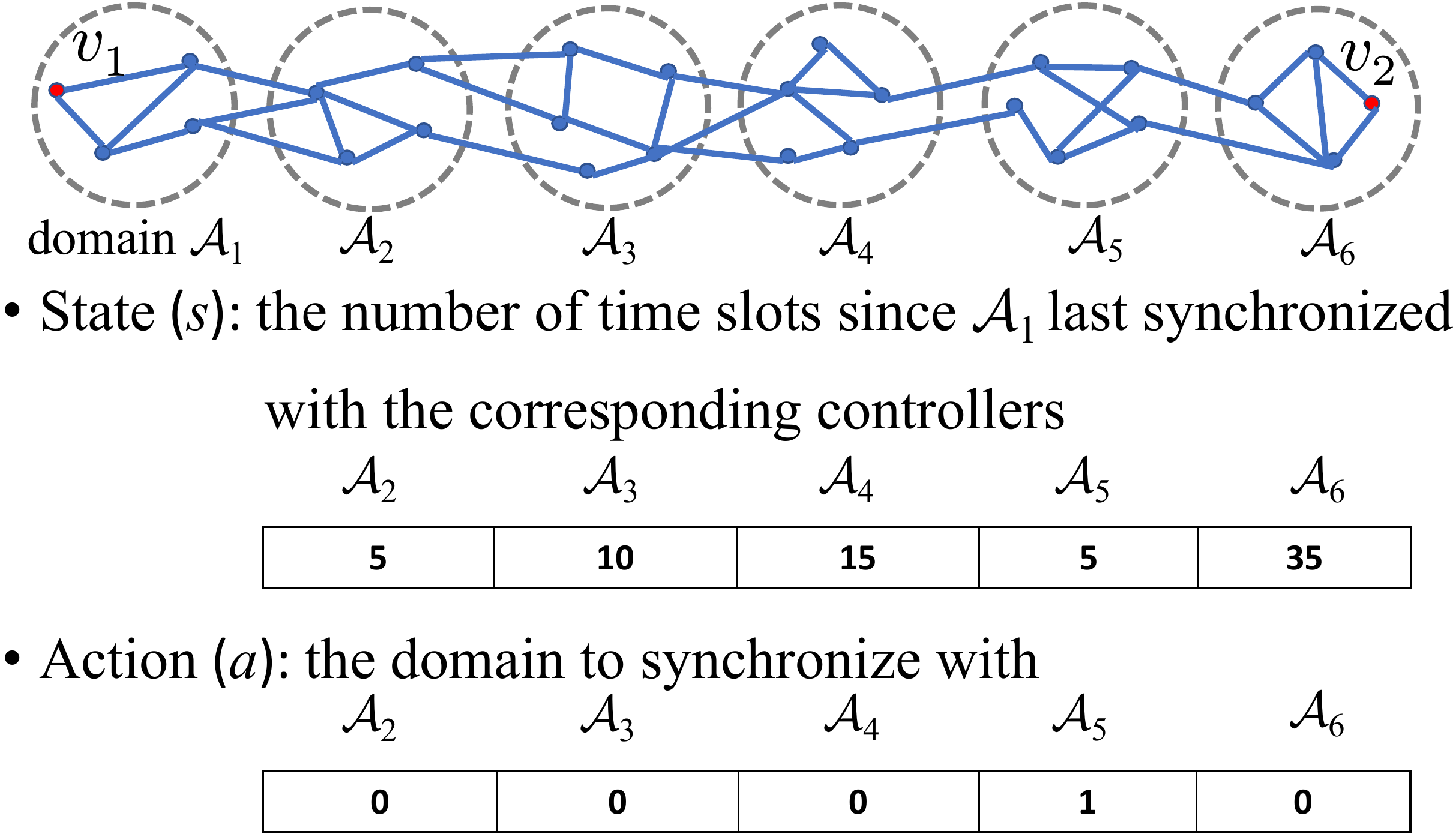}
	\caption{A state-action example for the MDP formulation.}
	\label{fig:state_action_example}
\end{figure}
The MDP formulation is demonstrated in an example in Fig.~\ref{fig:state_action_example}, where there are 6 domains on the domain-wise path between the source and destination nodes. The first entry in the state vector indicates that the last synchronization between the source controller and the controller of $\mathcal{A}_{2}$ took place 5 time slots ago. The action vector consists of binary entries where $1$ indicates that the source controller will synchronize with the corresponding domain at current time slot and $0$ the opposite. The action vector in the above example indicates that under the synchronization budget of $1$, the source controller will synchronize with  $\mathcal{A}_{5}$ only.

The \emph{optimal action} at each state is defined as the action that yields the maximum long-term reward, which is defined as the discounted sum of the expected immediate reward of all future state-action pairs from the current state. The reward for the state-action pair $\Delta t$ steps ahead of the current state is discounted by $\gamma^{\Delta t}$, where $\gamma$ is called the \emph{discount factor} and $0<\gamma<1$. Here, $\gamma$ trades off the importance between the current and the future reward. Therefore, starting from an initial state $s_0$, the problem is formulated to maximize the long-term accumulated reward expressed in the following Bellman equation by selecting a sequence of actions ${\{a_{t}\}}_{t=0}^{T}$:
\begin{equation}\label{eq:long_term_reward}
V(s_0) = \mathbb{E}\bigg[\sum_{t=0}^{T}\gamma^t R(s_{t},a_{t})|s_0\bigg],
\end{equation}
where $s_t$ and $a_t$ constitute the state-action pair at time $t$, and $T$ is the time horizon of the synchronization optimization problem. 

\section{Deep-Q (DQ) Scheduler}
\label{sec:dq_scheduler}
To solve the formulated MDP, we use RL techniques to find the sequence of actions that maximize the Bellman equation in (\ref{eq:long_term_reward}). For RL, imagine an agent who jumps from state to state in the formulated MDP by taking some actions associated with certain rewards. The agent's goal is to discover a sequence of state-action pairs, called a \emph{policy}, that maximize the accumulated time-discounted rewards. By interacting with the MDP, the agent's experiences build up which finally lead to the discovery of the \emph{optimal} policy. For this problem, one important aspect is how the agent memorizes its experiences. Traditionally, the storage of experiences in tabular fashion is used. However, this approach is impractical in many RL tasks because of the lack of generalization for large state-action space. Indeed, the state-action space is enormous in our controller synchronization problem. Consider the example in Fig.\ref{fig:state_action_example} assuming the time horizon is $300$ time slots and the synchronization budget is 1, there are as many as $300^{5}$ states and $5$ actions associated with each state. In light of this, function approximators \cite{bertsekas1995neuro} have been proposed, among which Deep Neural Network (DNN) \cite{hagan1996neural} is a suitable candidate which finds its successes in many recent applications~\cite{mnih2015human}~\cite{silver2016mastering}. Motivated by this, we therefore use DNN as the value function approximator in our DQ Scheduler, see Section~\ref{DDN} for details. Finally, we present the training algorithm for DQ Scheduler in Section~\ref{dq_scheduler}. 

\subsection{Q-learning with Parameterized Value Function}\label{q_learning}
\emph{$Q$-learning} \cite{watkins1992q} is a classic RL algorithm with performance guarantees under certain conditions \cite{kearns1999finite}. $Q$- learning uses the \emph{$Q$-function} to estimate the quality of a state-action pair:
$$
\mathcal{S} \times \mathcal{A} \rightarrow \mathbb{R}.
$$
\smallskip
In particular, the optimal $Q$-function for a state-action pair in $Q$-placement is defined as:
\begin{equation}
\label{eq:iterative_update}
Q^{*}(s,a) = \mathbb{E}[R(s,a)+\gamma\max_{a^{\prime} \in A_{s^{\prime}}} Q^{*}(s^{\prime},a^{\prime})],
\end{equation}
where $s^{\prime}$ and $a^{\prime}$ is the state-action pair at the next time slot, $A_{s^{\prime}}$ is the set of actions available at the next state $s^{\prime}$. 
Since we use DNN as the function approximator of the agent's $Q$-function $Q(s,a)$, it is parameterized by the set of adjustable parameters $\theta$ representing the weights of the DNN. The parameterized $Q$-function and optimal $Q$-function are denoted by $Q_{\theta}(s,a)$ and $Q^*_{\theta}(s,a)$, respectively.  The \emph{value iteration update}  \cite{watkins1992q} of the $Q$-function is based on (\ref{eq:iterative_update}), which uses the best estimation of the future reward of the next state to update current $Q$-function, thus approximating the optimal $Q^*_{\theta}(s,a)$. During the update, $\theta$ is adjusted to reduce the gap between the estimated and the optimal values. In particular, the following loss function using the mean-squared error measurement is defined for adjusting $\theta$:  \looseness = -1
\begin{equation}\label{eq:Q_update}
\begin{split}
L(\theta) & =\mathbb{E}[\big(y-Q_{\theta}(s,a)\big)^{2}],\\
\end{split}
\end{equation}
where 
\begin{equation}\label{eq:target}
y = R(s,a)+\gamma\max_{a^{\prime} \in A_{s^{\prime}}}Q_{\theta}(s^{\prime},a^{\prime})
\end{equation}
is the estimation of the maximum accumulated future reward. 

Then, by differentiating $L(\theta)$ w.r.t. $\theta$, we have the following gradient:
\begin{multline}
\nabla_{\theta}L(\theta)  =-2\mathbb{E}[\big(R(s,a)+\gamma\max_{a^{\prime} \in A_{s^{\prime}}}Q_{\theta}(s^{\prime},a^{\prime})\\-Q_{\theta}(s,a)\big)\nabla_{\theta}Q_{\theta}(s,a)].
\end{multline}
Then, weights of the DNN are updated for the next iteration: \looseness = -1
\begin{equation}
\theta \leftarrow \theta - \alpha\nabla_{\theta}L(\theta),
\end{equation}
where $\alpha$ is the step size. 
Note that the gradient descent update iterations of $\theta$ is different from canonical supervised learning because the training target $y = R(s,a)+\gamma\max_{a^{\prime} \in A_{s^{\prime}}}Q_{\theta}(s^{\prime},a^{\prime})$ is generated by the same parameterized $Q$-function $Q_{\theta}(s,a)$ that is being trained.  Therefore, to improve stability and performance of the training process, we improve the training algorithm in the following ways. 

\emph{\textbf{1)}}. We maintain a delayed version of the $Q$-function, $Q_{\theta^{'}}(s,a)$, for the estimation of the maximum next state reward, which was proposed~\cite{mnih2015human} to improve the stability of their DQN for playing Atari games. As such, the target function in (\ref{eq:target}) is updated to 
\begin{equation}\label{eq:target_updated}
y = R(s,a)+\gamma\max_{a^{\prime} \in A_{s^{\prime}}}Q_{\theta^{'}}(s^{\prime},a^{\prime}).
\end{equation}
The delayed $Q$-function is updated with the newest weights every $C$ steps by setting $\theta^{'} = \theta$. 

\emph{\textbf{2)}}. To overcome the overestimation of action values, we implement \emph{Double Q-learning}\cite{hasselt2010double} to address the positive bias in estimation introduced when the maximum expected action values are instead approximated by the maximum action values in $Q$-learning. Specifically, we use the up-to-date $Q$-function $Q_{\theta}(s',a')$ to determine $a'^{*} = \arg\max_{a'}Q_{\theta}(s',a')$, and the accumulated reward of the returned action $a'^{*}$ is estimated by the delayed $Q$-function using (\ref{eq:target_updated}). 

\emph{\textbf{3)}}. We implement the ``replay memory" \cite{lin1993reinforcement} in which some of the agent's past experiences in $(s,a,r)$ tuples are stored and maybe used more than once for training. In particular, a matrix $\mathcal{D}$ is created that can store up to $N$ $(s,a,r)$ tuples. At each training iteration where $Q$-learning update takes place, samples of experiences are pulled randomly from $\mathcal{D}$ for training. 
%
%

\subsection{The Design of the Deep Neural Network (DDN)}\label{DDN}
The Parameterized $Q$-function is implemented by a \emph{Multilayer Perceptron (MLP)} \cite{kumar2011multilayer} consisting of input/output and three hidden layers. Let $m$ denote the number of domains on the domain-wise path. The input to the MLP is of dimension $2(m-1)\times 1$. The first $m-1$ entries store the state of the MDP, and the rest $m-1$ binary entries store the action. The output of the MLP is the maximum predicted accumulated time-discounted reward given the state-action input. The three hidden layers consist of 128, 64, and 32 hidden neurons, respectively. The MLP is realized using Keras \cite{chollet2018keras} model with TensorFlow\cite{abadi2016tensorflow}, in which Adam is chosen as the optimizer and Rectified Linear Unit (ReLU)\cite{nair2010rectified} is employed as activation functions for all neurons except for the output layer. \looseness = -1

%
%

\subsection{The Training Algorithm}\label{dq_scheduler}
To train the DNN that represents the parameterized $Q$-function, we need to first initialize the matrix $\mathcal{D}$, i.e., the agent's ``reply memory". This is different from traditional online $Q$-learning where the update of $Q$-matrix relies only on the current $(s,a,r)$ tuple.  Instead,  history data are used for the training of the DNN for stability reasons. In particular, $\mathcal{D}$ is initially filled up with several $(s,a,r)$ tuples for training. As time proceeds, more $(s,a,r)$ tuples are generated and recorded in matrix $\mathcal{D}$ for training. Since we limit the number of entries of $\mathcal{D}$ to $N$, old $(s,a,r)$ tuples are gradually replaced by new entries; see Section~\ref{sec:evaluation} for the simulation settings that generate the training data. There are many ways in which actions can be selected. Traditionally, $\epsilon$-greedy algorithm\cite{Thrun92efficientexploration} is used in $Q$-learning; there are also new exploration strategies proposed which are tailored for the DNN settings, such as bootstrapped DQN\cite{osband2016deep} and UCB\cite{chen2017ucb}. According to~\cite{azar2012sample}, the exploration strategy that generates all state-action pairs uniformly at random is better for training. Therefore, our training algorithm takes in data generated by random exploration. The training process is summarized in Algorithm~\ref{alg:aglorithm1}. After the training algorithm terminates, the returned parameterized $Q$-function will be able to estimate the best action to take at each state, thus approximating the optimal.  

\begin{algorithm}[tb] 
	\small
	\SetKwInOut{Input}{input}\SetKwInOut{Output}{output}
	\Input{DNN model settings; distributed SDN settings; simulation program for generating rewards; delay $C$ of the delayed $Q$-function.}
	\Output{Trained parameterized $Q$-function.}
	Initialize the $Q$-function $Q_{\theta}(s,a)$ by instantiating the DNN with random initial weights and biases settings;\\
	Initialize matrix $\mathcal{D}$ with history $(s,a,r)$ tuples; \\
	Initialize the delayed $Q$-function $Q_{\theta^{'}}(s,a) = Q_{\theta}(s,a)$ ; \\
	Set initial state $s_{0}$; $t = 0$;\\\
	\While{ $t\leq T$}
	{		
		\ForEach {time instant $t$}
		{ Select an action $a_{t}$ randomly;\\
			Pass on the $(s_{t},a_{t})$ to the simulation program and get return $r_{t}$;\\
			Store $(s_{t},a_{t},r_{t})$ in $\mathcal{D}$;\\
			Pull random minibatch $\mathcal{D}_{t}$ of $(s_{i},a_{i},r_{i})$ from $\mathcal{D}$;
		   
		   \ForEach {$(s,a,r)$ in $\mathcal{D}_{t}$ }
			{ 
				$a^{*}_{j+1} = \arg\max_{a_{j+1}}Q_{\theta}(s_{j+1},a_{j+1})$;\\
				$y_{j} =r_{j}+\gamma Q_{\theta^{'}}(s_{j+1},a^{*}_{j+1})$;\\
				Calculate the gradient: $\nabla_{\theta}L(\theta)  =-2(y_{j}-Q_{\theta}(s_{j},a_{j}))\nabla_{\theta}Q_{\theta}(s_{j},a_{j})$;\\
				Update weights: $\theta \leftarrow \theta - \alpha\nabla_{\theta}L(\theta)$;\\
			}
			
			\If{$t\mod C = 0$}
			{
				$Q_{\theta^{'}}(s,a) = Q_{\theta}(s,a)$.
			}
		}
	}\caption{Training algorithm for DQ Scheduler}\label{alg:aglorithm1}
\end{algorithm}
\normalsize

\section{Evaluation}
\label{sec:evaluation}
In this section we present the performance evaluation of the proposed DQ Scheduler comparing to other two default controller synchronization schemes. Specifically, we first introduce the network and its dynamicity model used for building simulation networks in Section~\ref{net_dynamic_model}. Then, the settings and datasets used are described in Section~\ref{net_settings}. Finally, we present the evaluation results and analysis in Section~\ref{evaluation_results}. \looseness = -1

\subsection{Network and Dynamicty Model}
\label{net_dynamic_model}
%

\emph{\textbf{Network Model}}: The topology of domain $i$ with $n_{i}$ nodes is modeled as an undirected graph, where $n_{i}$ nodes are connected following a given \emph{intra-domain degree distribution}, i.e.,  the distribution of the number of neighbouring nodes of an arbitrary node. 
Then for any two neighbouring domains $\mathcal{A}_i$ and $\mathcal{A}_j$, we (i) randomly select two nodes $w_1$ from $\mathcal{A}_i$ and $w_2$ from $\mathcal{A}_j$ and connect these two nodes if link $w_1w_2$ does not exist, and (ii) repeat such link construction process between $\mathcal{A}_i$ and $\mathcal{A}_j$ $\beta_{i,j}$ times. By this link construction process, network topology $\mathcal{G}=(V,E)$ is therefore formed ($V/E$: set of nodes/links in $\mathcal{G}$, $|V|=\sum_{i=1}^{q}n_{i}$, where $q$ is  the number of domains). 

\emph{\textbf{Dynamicity of the network}}: We model the dynamic pattern of intra-domain topologies using a simple edge rewire model. Specifically, at each time slot, $e_{i}$ new edges are added randomly between nodes in domain $i$ before $e_{i}$ existing edges are randomly selected and removed. If this random edge rewire procedure results in a fragmented network topology, we randomly add the minimum number of edges to connect all components to make it a connected graph again. The edge weights of all newly added edges are generated by the given edge weight distribution. It should be noted that our DQ Scheduler does not have any requirement on the dynamicity model of the network, the rewire model we propose only serves as a simple and representative example. 


\begin{figure*}
	\smallskip
	\centering
	\begin{subfigure}[b]{0.32\textwidth}
		\centering
		\includegraphics[width=\linewidth,height=0.7\linewidth]{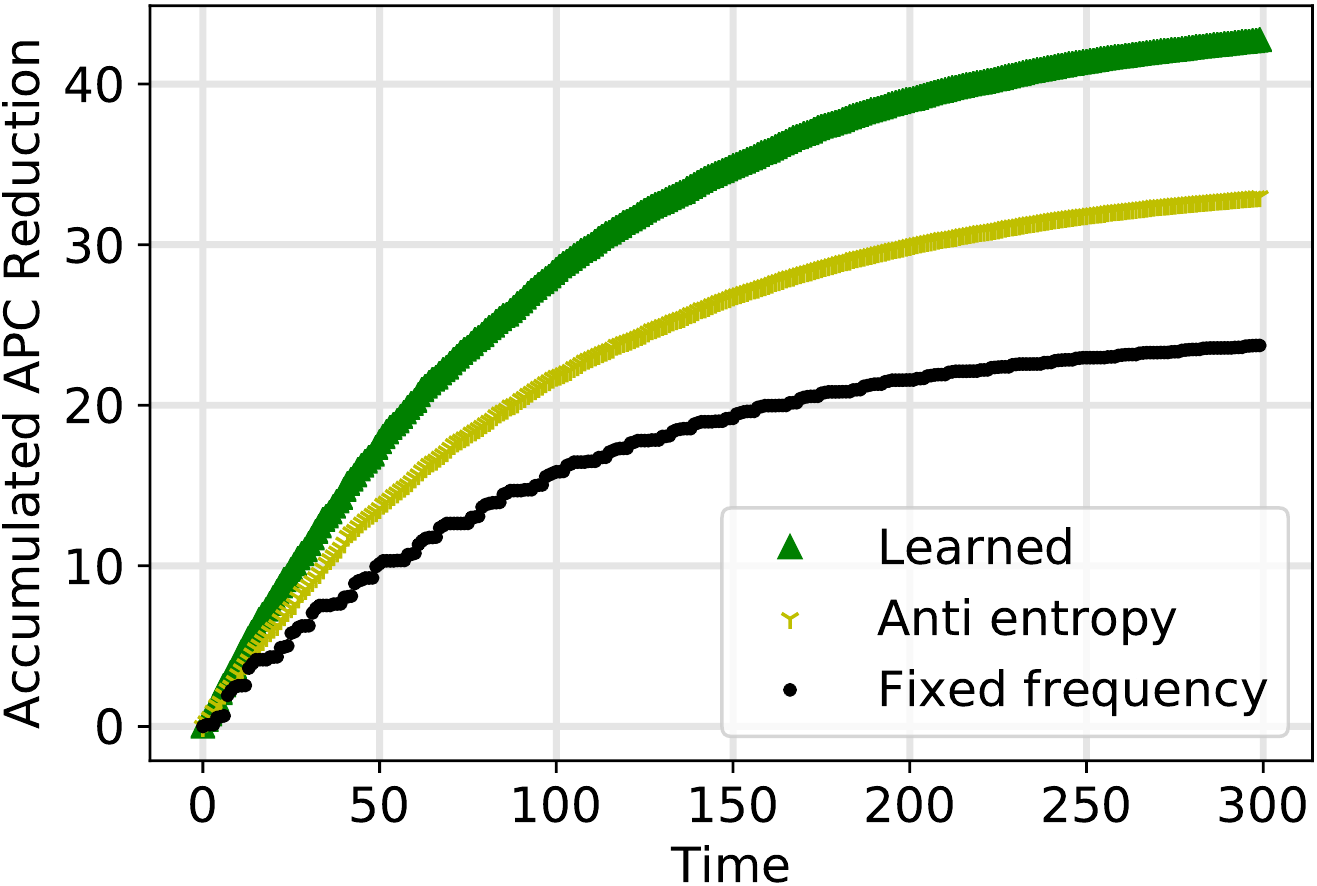}
		\caption{Accumulated APC reduction, $m=6$.}
		\label{fig:s1}
	\end{subfigure}
	\begin{subfigure}[b]{0.32\textwidth}
		\centering
		\includegraphics[width=\linewidth,height=0.7\linewidth]{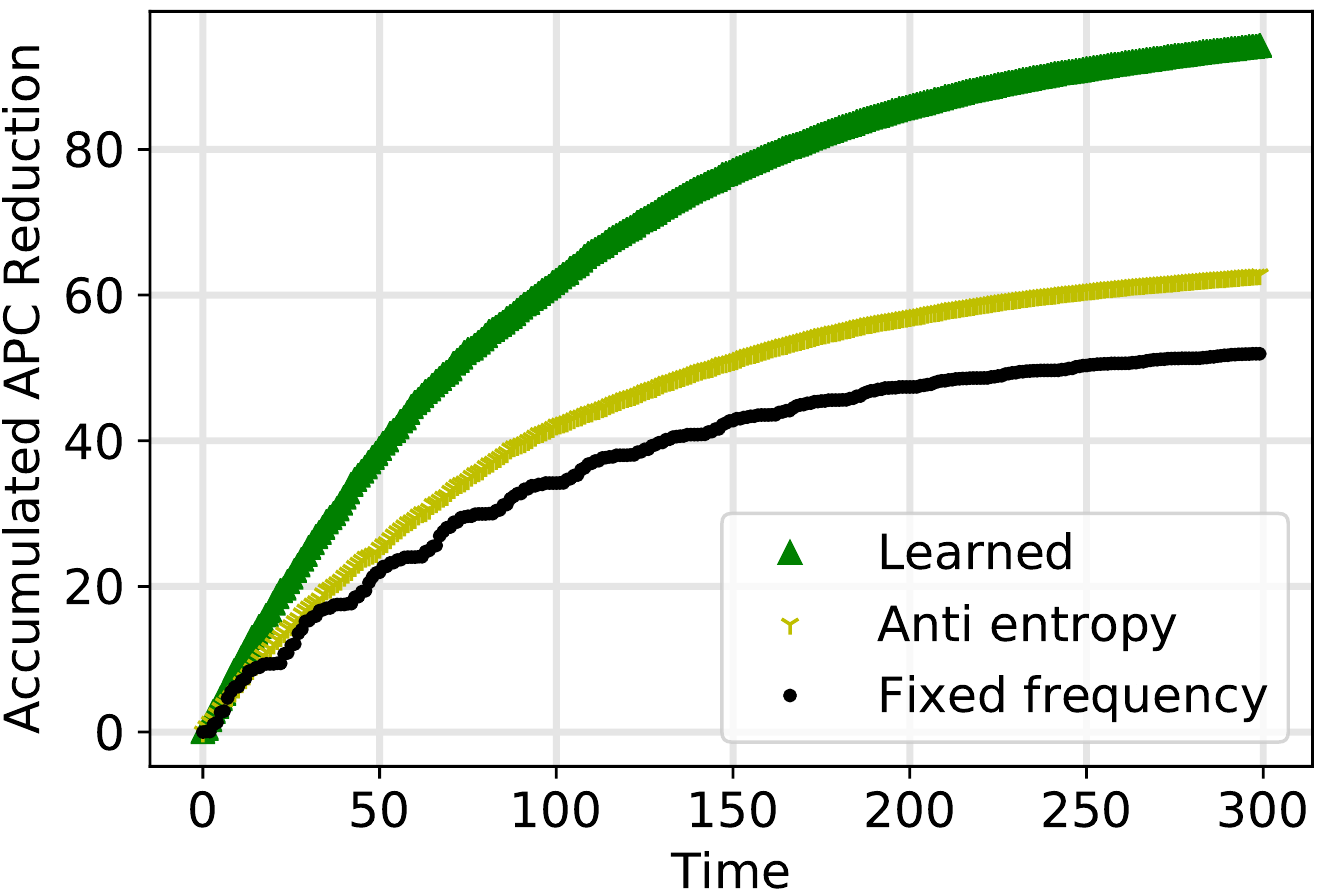}
		\caption{Accumulated APC reduction, $m=10.$}
		\label{fig:s2}
	\end{subfigure}
	\begin{subfigure}[b]{0.32\textwidth}
		\centering
		\includegraphics[width=\linewidth,height=0.7\linewidth]{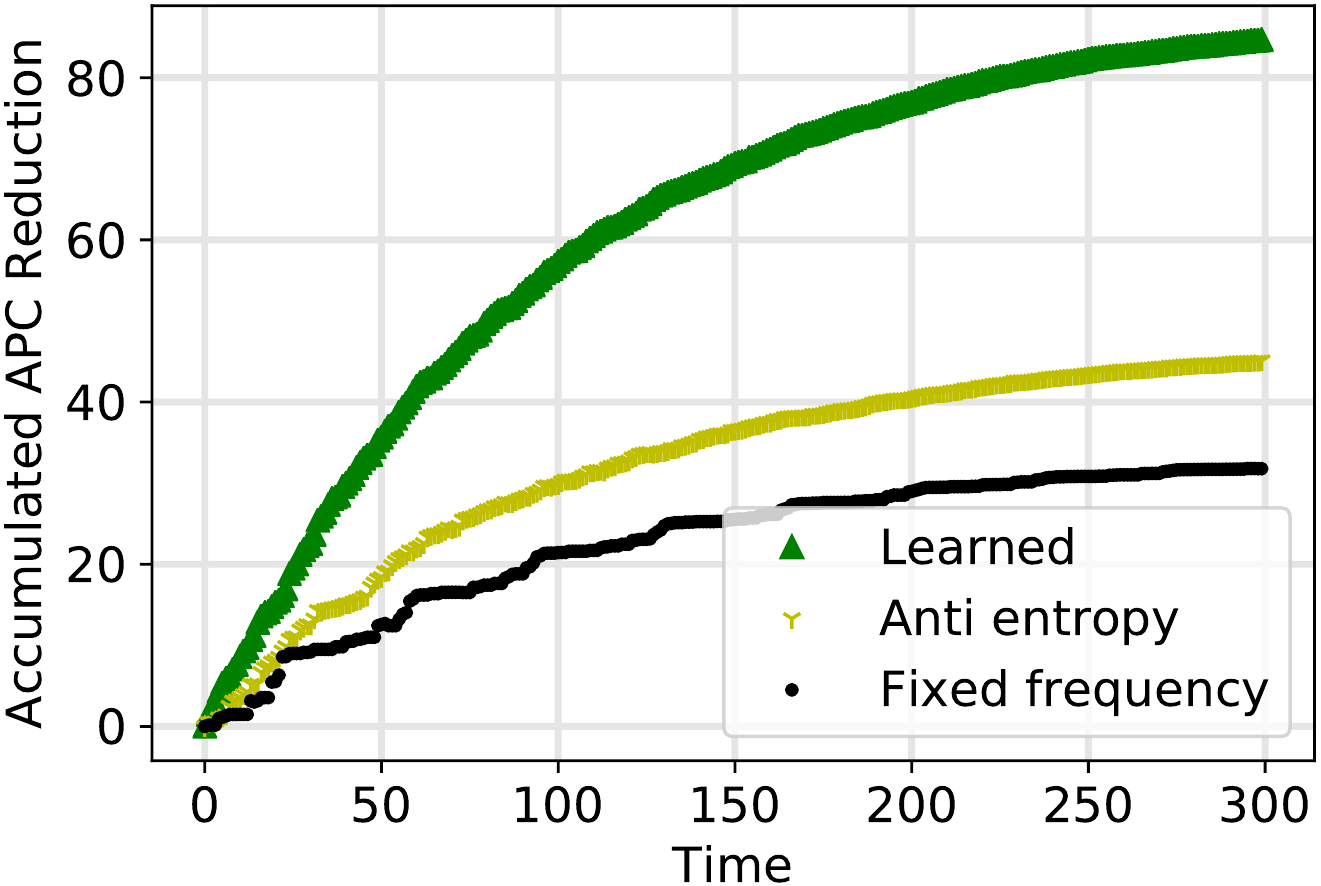}
		\caption{Accumulated APC reduction, $m=12$.}
		\label{fig:s3}
	\end{subfigure}   
	\begin{subfigure}[b]{0.32\textwidth}
		\centering
		\includegraphics[width=\linewidth,height=0.7\linewidth]{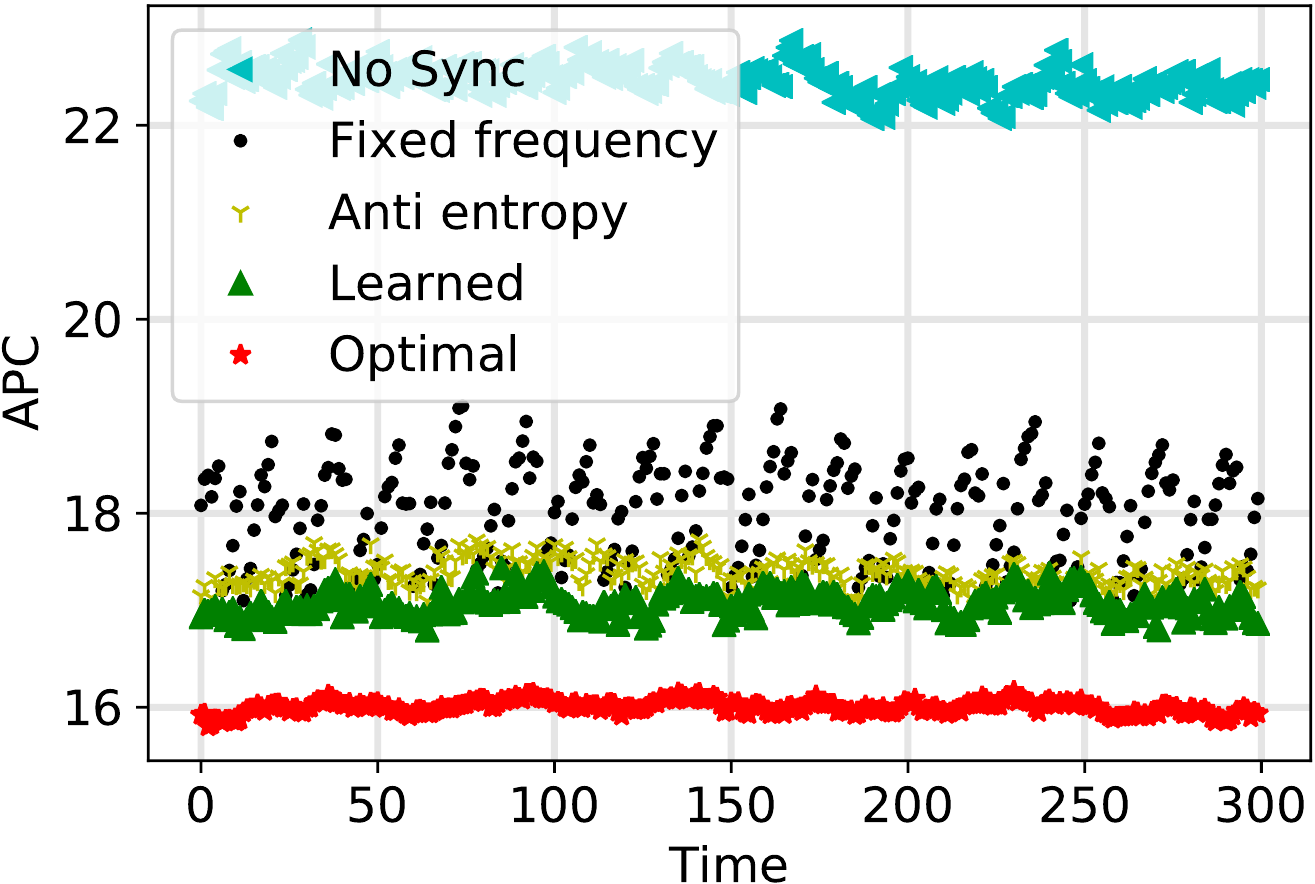}
		\caption{Immediate APC, $m=6$.}
		\label{fig:s4}
	\end{subfigure}
	\begin{subfigure}[b]{0.32\textwidth}
		\centering
		\includegraphics[width=\linewidth,height=0.7\linewidth]{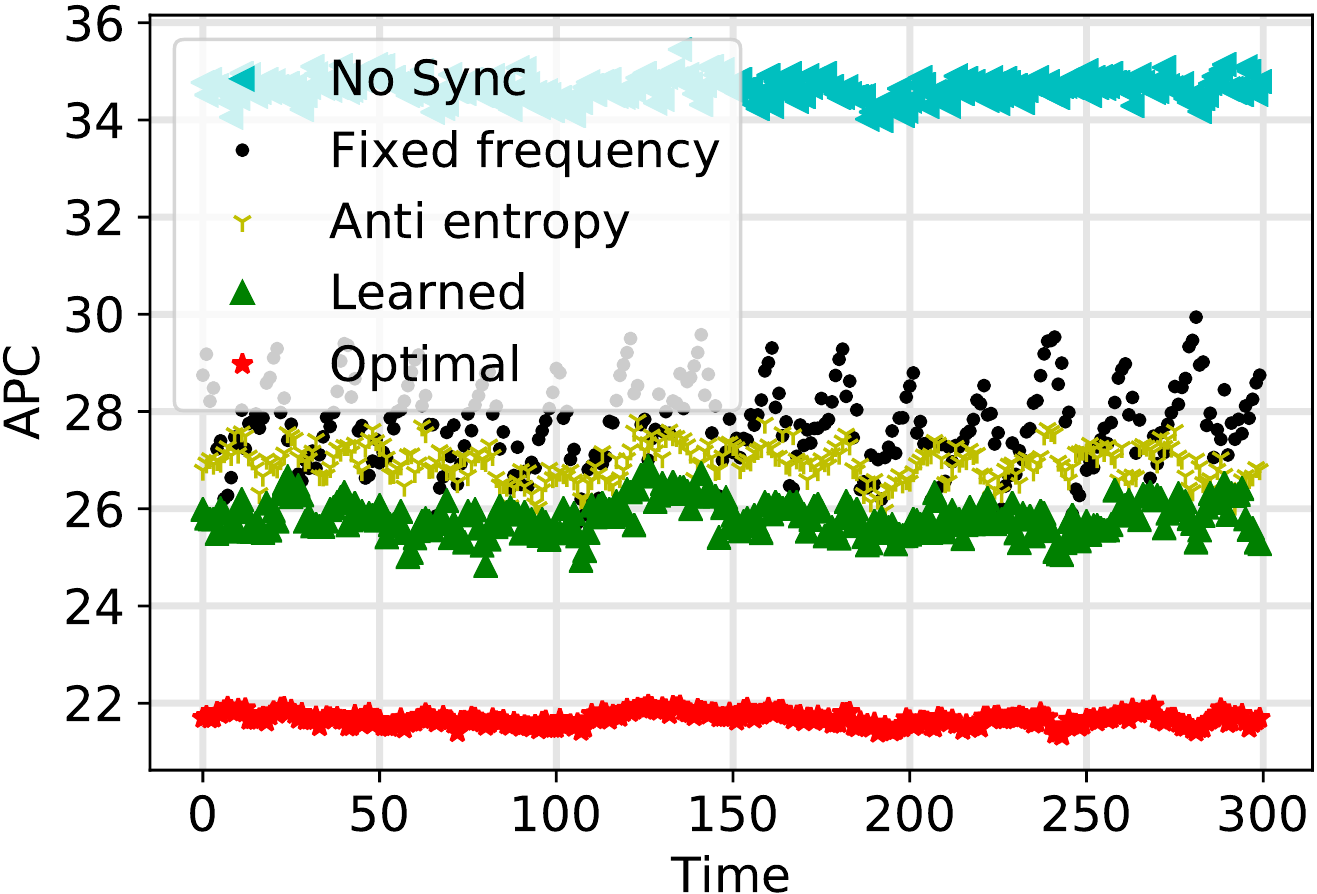}
		\caption{Immediate APC, $m=10$.}
		\label{fig:s5}
	\end{subfigure}    
	\begin{subfigure}[b]{0.32\textwidth}
		\centering
		\includegraphics[width=\linewidth,height=0.7\linewidth]{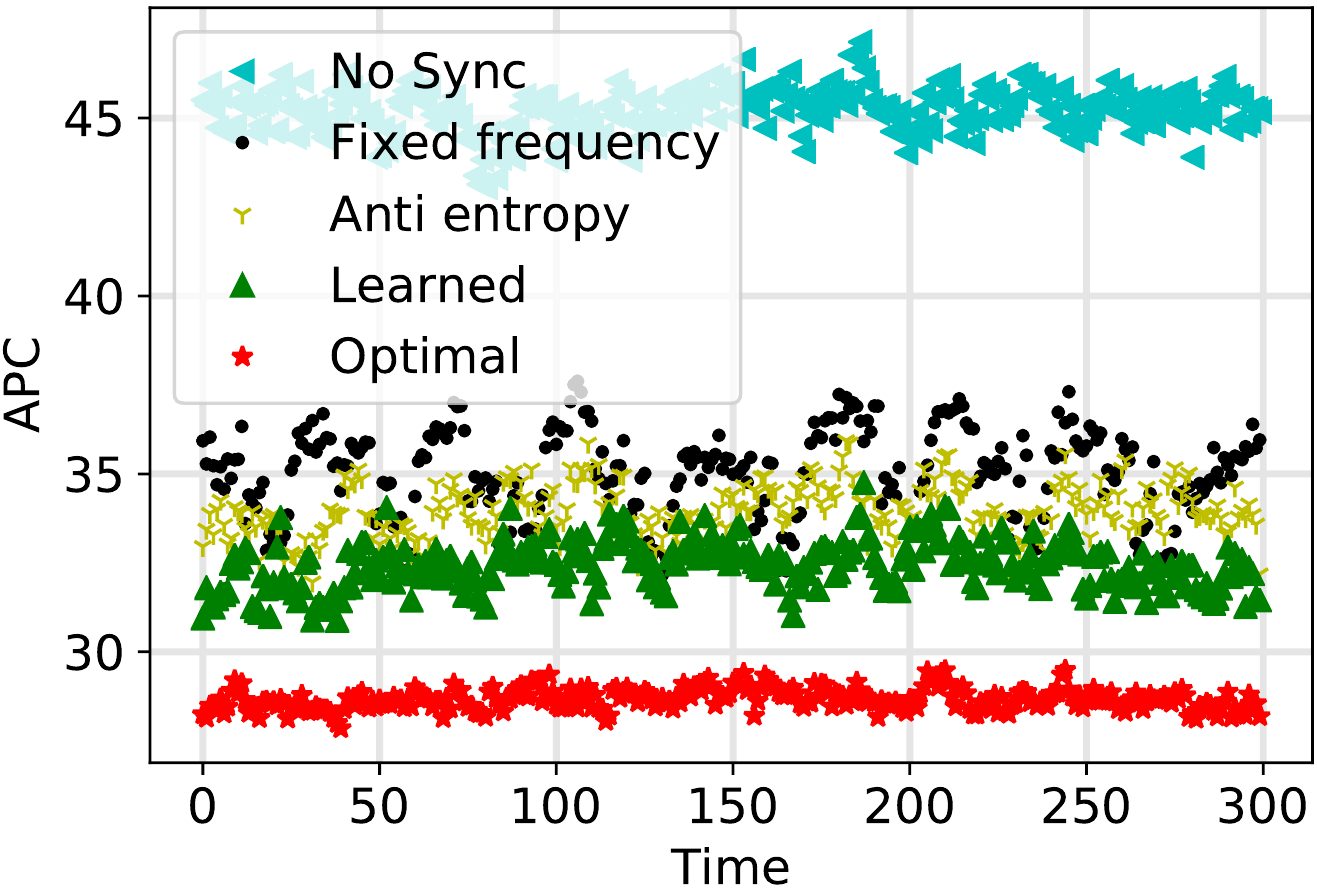}
		\caption{Immediate APC, $m=12$.}
		\label{fig:s6}
	\end{subfigure}  
	\caption{Evaluation results.}
	\label{fig:simulation_results}   
\end{figure*}
\subsection{Network Settings}
\label{net_settings}

According to the path construction mechanism described in Section~\ref{path_construction}, the domain-wise paths extracted w.r.t. source/destination node pairs are always in a linear fashion such as  the example in Fig.~\ref{fig:state_action_example}. Let $m$ be the number of domains on the domain-wise path with the indices of domains sequentially labelled from $1$ (source domain) to $m$ (destination domain). Our evaluation considers three scenarios where  $m=6$, $m=10$, and $m=12$. The degree distributions used is extracted from the RocketFuel Project\cite{Rocketfuel}, where we use data from``AS1239". The weights of edges are randomly drawn from the set $\{1,2,5,8,10,12\}$ with the corresponding probability being $10\%,10\%,10\%,30\%,20\%,$ and  $20\%$, respectively. In addition, the synchronization budget is set to $1$ for all evaluations. Let $\mathcal{N}_{m} = [n_{1},n_{2},\dots,n_{m}]$, and $\mathcal{E}_{m} = [e_{1},e_{2},\dots,e_{m}]$  be the vectors of the number of nodes, and the number of edge rewires at each time slot for the $m$ domains on the domain-wise path, respectively. In addition, $\mathcal{B}_{m} = [\beta_{1,2},\beta_{2,3},\dots,\beta_{m-1,m}]$ is the vector of the gateway connection parameters between all pairs of directly connected domains on the domain-wise path with $m$ domains. The parameters used for three evaluation scenarios are listed in Table~\ref{tb:parameters}. Note that DQ Scheduler \emph{does not} have any requirements for aforementioned parameters.

\begin{table}
	\centering
	\caption{Evaluation Parameters}
	\begin{tabular}{c| M{0.25\linewidth} |M{0.25
				\linewidth}|M{0.25\linewidth}  M{0.25\linewidth} }
		\hline
		$m$ & $\mathcal{N}_{m}$&$\mathcal{E}_{m}$ & $\mathcal{B}_{m}$ \\ \hline
		$6$ &\makecell{ $[100,300,550,$\\$150,210,420]$}& \makecell{$[5,1,40,60,2,120]$}& \makecell{$[5,20,8,20,12]$} \\ \hline
		$10$ & \makecell{$[100,300,550,150,$\\$210,420,380,$\\$520,120,340]$} & \makecell{$[5,1,160,1,180,$\\$150,40,150,1,40]$} & \makecell{$[5,30,50,20,$\\$15,20,20,30,10]$} \\ \hline
		$12$ & \makecell{$[130,50,550,150,$\\$80,420,380,330,$\\$250,150,80,100]$}  & \makecell{$[10,1,100,1,1,120,$\\$1,180,130,1,1,1]$}& \makecell{$[12,30,25,5,15,$\\$10,20,5,8,15,20]$} \\
		\hline
	\end{tabular}\label{tb:parameters}
\end{table}

The anti-entropy \cite{muqaddas2016inter} and fixed-frequency synchronization \cite{aslan2016adaptive} schemes are employed as benchmarks for evaluating the performance of DQ Scheduler.  The core of anti-entropy protocol, which is implemented in the ONOS controller, is based on a simple gossip algorithm that each controller randomly chooses another controller to synchronize. In comparison, the fixed-frequency algorithms synchronize controller pairs at constant rates, which may vary for different controller pairs subject to various objectives. For this evaluation, all controllers synchronize at the same rate.  

\subsection{Evaluation Results}
\label{evaluation_results}

The evaluation results of the DQ Scheduler for three scenarios are presented in Fig.\ref{fig:simulation_results}, where Fig.\ref{fig:s1}--Fig.\ref{fig:s3} show the performance in terms of the objective stated in Section~\ref{sec:objective}. Fig.\ref{fig:s4}--Fig.\ref{fig:s6} show the APC of packets delivered under three controller synchronization schemes. 

\emph{1) Superiority of DQ Scheduler for long-term routing quality:} Recall that our objective is aimed at overall routing quality over a period of time, i.e., to maximize $V = \mathbb{E}\bigg[\sum_{t=0}^{T}\gamma^t R(s_{t},a_{t})\bigg]$. The evaluation results in Fig.\ref{fig:s1}--Fig.\ref{fig:s3} confirm the superiority of DQ Scheduler in achieving this goal. In particular, during the testing period of 300 time slots, DQ scheduler outperforms the anti-entropy algorithm by $31.2\%$, $58.3\%$, and $95.2\%$; the algorithm with constant synchronization rate by $90.9\%$, $90\%$, and $173.3\%$, for three scenarios, respectively. 

\emph{2) Superiority of DQ Scheduler for immediate routing quality:} Although DQ Scheduler is trained to maximize the accumulated APC reduction over time, surprisingly, its synchronization decisions also lead to the lowest APC in real-time among three algorithms tested, as shown in Fig.\ref{fig:s4}--Fig.\ref{fig:s6}. This means that the DQ scheduler optimizes the accumulated APC reduction in a way that the immediate and long-term performance are balanced, since there is not a period in which the immediate performance is worsened for better future performance according to these results. In these evaluations,  the APCs are also compared to the ``optimal" case where the source domain is always synchronized with all other domains on the domain-wise path,s and to the ``worst" case when there is no synchronization (``no sync") between any controllers. 

\emph{3) Other findings:} Compared to the other benchmark algorithms, DQ Scheduler's performance is more stable when the domain-wise path involves more domains. In contrast, two benchmark algorithms' performance first improves and then becomes worse when the number of domains on the domain-wise path increases from $6$ to $10$, and then to $12$. In addition, we realize that the performance degradation of no controller synchronization is more concerning when there are more domains involved on the domain-wise path, as ``no sync" performance are worsened by $37.5\%$, $59.1\%$, and $61\%$ in three scenarios, comparing to the optimal cases. This highlights the important role of controller synchronizations.  \looseness = -1

\section{Related Work}
\label{sec:related_work}
\subsubsection{Distributed SDN}
Many research efforts are directed to the design of distributed SDN  controller architecture. Specifically,  OpenDaylight~\cite{opendaylight} and ONOS~\cite{berde2014onos} are two state-of-the-art SDN controllers proposed to realize logically centralized but physically distributed SDN architecture. In addition, controllers such as Devoflow~\cite{curtis2011devoflow} and Kandoo~\cite{hassas2012kandoo} are designed with their specific aims. However, most of these controller architectures do not emphasize or justify detailed controller synchronization protocols they employ. 

\subsubsection{Controller Synchronizations}
Most existing works on controller synchronization assume either strong or eventual consistency models\cite{sakic2018impact}, for which our work uses the latter. The authors in \cite{levin2012logically} show that certain network applications can rely on the eventual consistency to deliver acceptable performance. This work \cite{guo2014improving} shows how to avoid network anomalies such as forwarding loops and black holes under the eventual consistency assumption. Similar to our approach, the works \cite{aslan2016adaptive}\cite{sakic2017towards} propose dynamic adaptation of synchronization rate among controllers. Compared to these works, DQ Scheduler is much more versatile in that there is no assumptions on the network and the policy learning process is automated given any network conditions.   

\subsubsection{Reinforcement Learning in SDN}
Some recent high-profile successes~\cite{mnih2015human}~\cite{silver2016mastering} attract enormous interests in applying RL techniques to solve complicated decision making problems. In the context of SDN, the authors in \cite{zhang2018q} apply RL-based algorithms to solve service placement problem on SDN switches. This work\cite{lin2016qos} also discusses the routing problem in SDN using RL techniques. However, the discussion is only limited to intra-domain routing under strong assumptions on the network topology. In addition, tabular settings are used in this work without generalizations. 
\section{Conclusion}
\label{sec:conclusion}
In this paper, we investigated the controller synchronization problem with limited synchronization budget in distributed SDN, for which our aim was to find the policy that maximizes the benefits of controller synchronizations over a period of time. We formulated the controller synchronization problem as an MDP problem. An RL-based algorithm which uses the DNN to represent its value function, called the DQ Scheduler, was proposed to solve the formulated MDP.  Evaluation results showed that DQ Scheduler offers almost twofold performance improvement comparing to the state-of-the-art SDN controller synchronization solutions. 

\bibliographystyle{IEEEtran}
\bibliography{reference}

\begin{thebibliography}{10}
\providecommand{\url}[1]{#1}
\csname url@samestyle\endcsname
\providecommand{\newblock}{\relax}
\providecommand{\bibinfo}[2]{#2}
\providecommand{\BIBentrySTDinterwordspacing}{\spaceskip=0pt\relax}
\providecommand{\BIBentryALTinterwordstretchfactor}{4}
\providecommand{\BIBentryALTinterwordspacing}{\spaceskip=\fontdimen2\font plus
\BIBentryALTinterwordstretchfactor\fontdimen3\font minus
  \fontdimen4\font\relax}
\providecommand{\BIBforeignlanguage}[2]{{%
\expandafter\ifx\csname l@#1\endcsname\relax
\typeout{** WARNING: IEEEtran.bst: No hyphenation pattern has been}%
\typeout{** loaded for the language `#1'. Using the pattern for}%
\typeout{** the default language instead.}%
\else
\language=\csname l@#1\endcsname
\fi
#2}}
\providecommand{\BIBdecl}{\relax}
\BIBdecl

\bibitem{kreutz2015software}
D.~Kreutz, F.~M. Ramos, P.~E. Verissimo, C.~E. Rothenberg, S.~Azodolmolky, and
  S.~Uhlig, ``Software-defined networking: A comprehensive survey,''
  \emph{Proceedings of the IEEE}, vol. 103, no.~1, pp. 14--76, 2015.

\bibitem{panda2013cap}
A.~Panda, C.~Scott, A.~Ghodsi, T.~Koponen, and S.~Shenker, ``{CAP} for
  networks,'' in \emph{ACM HotSDN}, 2013.

\bibitem{levin2012logically}
D.~Levin, A.~Wundsam, B.~Heller, N.~Handigol, and A.~Feldmann, ``Logically
  centralized?: state distribution trade-offs in software defined networks,''
  in \emph{ACM HotSDN}, 2012.

\bibitem{berde2014onos}
P.~Berde, M.~Gerola, J.~Hart, Y.~Higuchi, M.~Kobayashi, T.~Koide, B.~Lantz,
  B.~O'Connor, P.~Radoslavov, W.~Snow \emph{et~al.}, ``{ONOS}: {T}owards an
  open, distributed {SDN} {OS},'' in \emph{ACM HotSDN}, 2014.

\bibitem{muqaddas2016inter}
A.~S. Muqaddas, A.~Bianco, P.~Giaccone, and G.~Maier, ``Inter-controller
  traffic in {ONOS} clusters for {SDN} networks,'' in \emph{IEEE ICC}, 2016.

\bibitem{kotronis2015routing}
V.~Kotronis, A.~G{\"a}mperli, and X.~Dimitropoulos, ``Routing centralization
  across domains via sdn: A model and emulation framework for bgp evolution,''
  \emph{Computer Networks}, vol.~92, pp. 227--239, 2015.

\bibitem{bertsekas1995neuro}
D.~P. Bertsekas and J.~N. Tsitsiklis, ``Neuro-dynamic programming: an
  overview,'' in \emph{Proceedings of the 34th IEEE Conference on Decision and
  Control}, vol.~1.\hskip 1em plus 0.5em minus 0.4em\relax IEEE Publ.
  Piscataway, NJ, 1995, pp. 560--564.

\bibitem{hagan1996neural}
M.~T. Hagan, H.~B. Demuth, M.~H. Beale, and O.~De~Jes{\'u}s, \emph{Neural
  network design}.\hskip 1em plus 0.5em minus 0.4em\relax Pws Pub. Boston,
  1996, vol.~20.

\bibitem{mnih2015human}
V.~Mnih, K.~Kavukcuoglu, D.~Silver, A.~A. Rusu, J.~Veness, M.~G. Bellemare,
  A.~Graves, M.~Riedmiller, A.~K. Fidjeland, G.~Ostrovski \emph{et~al.},
  ``Human-level control through deep reinforcement learning,'' \emph{Nature},
  vol. 518, no. 7540, p. 529, 2015.

\bibitem{silver2016mastering}
D.~Silver, A.~Huang, C.~J. Maddison, A.~Guez, L.~Sifre, G.~Van Den~Driessche,
  J.~Schrittwieser, I.~Antonoglou, V.~Panneershelvam, M.~Lanctot \emph{et~al.},
  ``Mastering the game of go with deep neural networks and tree search,''
  \emph{nature}, vol. 529, no. 7587, p. 484, 2016.

\bibitem{watkins1992q}
C.~J. Watkins and P.~Dayan, ``Q-learning,'' \emph{Machine learning}, vol.~8,
  no. 3-4, pp. 279--292, 1992.

\bibitem{kearns1999finite}
M.~J. Kearns and S.~P. Singh, ``Finite-sample convergence rates for
  {Q}-learning and indirect algorithms,'' in \emph{Advances in Neural
  Information Processing Systems}, 1999, pp. 996--1002.

\bibitem{hasselt2010double}
H.~V. Hasselt, ``Double {Q}-learning,'' in \emph{NIPS}, 2010.

\bibitem{lin1993reinforcement}
L.-J. Lin, ``Reinforcement learning for robots using neural networks,''
  Carnegie-Mellon Univ Pittsburgh PA School of Computer Science, Tech. Rep.,
  1993.

\bibitem{kumar2011multilayer}
M.~Kumar and N.~Yadav, ``Multilayer perceptrons and radial basis function
  neural network methods for the solution of differential equations: a
  survey,'' \emph{Computers \& Mathematics with Applications}, vol.~62, no.~10,
  pp. 3796--3811, 2011.

\bibitem{chollet2018keras}
F.~Chollet \emph{et~al.}, ``Keras: The python deep learning library,''
  \emph{Astrophysics Source Code Library}, 2018.

\bibitem{abadi2016tensorflow}
M.~Abadi, P.~Barham, J.~Chen, Z.~Chen, A.~Davis, J.~Dean, M.~Devin,
  S.~Ghemawat, G.~Irving, M.~Isard \emph{et~al.}, ``Tensorflow: a system for
  large-scale machine learning.'' in \emph{OSDI}, vol.~16, 2016, pp. 265--283.

\bibitem{nair2010rectified}
V.~Nair and G.~E. Hinton, ``Rectified linear units improve restricted boltzmann
  machines,'' in \emph{2010 The 27th ICML}.

\bibitem{Thrun92efficientexploration}
S.~B. Thrun, ``Efficient exploration in reinforcement learning,'' Tech. Rep.,
  1992.

\bibitem{osband2016deep}
I.~Osband, C.~Blundell, A.~Pritzel, and B.~Van~Roy, ``Deep exploration via
  bootstrapped dqn,'' in \emph{Advances in neural information processing
  systems}, 2016, pp. 4026--4034.

\bibitem{chen2017ucb}
R.~Y. Chen, S.~Sidor, P.~Abbeel, and J.~Schulman, ``Ucb exploration via
  q-ensembles,'' \emph{arXiv preprint arXiv:1706.01502}, 2017.

\bibitem{azar2012sample}
M.~G. Azar, R.~Munos, and B.~Kappen, ``On the sample complexity of
  reinforcement learning with a generative model,'' \emph{arXiv preprint
  arXiv:1206.6461}, 2012.

\bibitem{Rocketfuel}
\BIBentryALTinterwordspacing
``Rocketfuel: An {ISP} topology mapping engine,'' University of Washington,
  2002. [Online]. Available:
  \url{http://www.cs.washington.edu/research/networking/rocketfuel/interactive/}
\BIBentrySTDinterwordspacing

\bibitem{aslan2016adaptive}
M.~Aslan and A.~Matrawy, ``Adaptive consistency for distributed sdn
  controllers,'' in \emph{2016 17th IEEE International Telecommunications
  Network Strategy and Planning Symposium (Networks)}.

\bibitem{opendaylight}
\BIBentryALTinterwordspacing
{T}he {O}pendaylight {C}ontroller. [Online]. Available:
  \url{https://www.opendaylight.org}
\BIBentrySTDinterwordspacing

\bibitem{curtis2011devoflow}
A.~R. Curtis, J.~C. Mogul, J.~Tourrilhes, P.~Yalagandula, P.~Sharma, and
  S.~Banerjee, ``Devoflow: {S}caling flow management for high-performance
  networks,'' in \emph{ACM SIGCOMM}, 2011.

\bibitem{hassas2012kandoo}
S.~Hassas~Yeganeh and Y.~Ganjali, ``Kandoo: {A} framework for efficient and
  scalable offloading of control applications,'' in \emph{ACM HotSDN}, 2012.

\bibitem{sakic2018impact}
E.~Sakic and W.~Kellerer, ``Impact of adaptive consistency on distributed sdn
  applications: An empirical study,'' \emph{IEEE Journal on Selected Areas in
  Communications}, 2018.

\bibitem{guo2014improving}
Z.~Guo, M.~Su, Y.~Xu, Z.~Duan, L.~Wang, S.~Hui, and H.~J. Chao, ``Improving the
  performance of load balancing in software-defined networks through load
  variance-based synchronization,'' \emph{Computer Networks}, vol.~68, pp.
  95--109, 2014.

\bibitem{sakic2017towards}
E.~Sakic, F.~Sardis, J.~W. Guck, and W.~Kellerer, ``Towards adaptive state
  consistency in distributed sdn control plane,'' in \emph{IEEE ICC}, 2017.

\bibitem{zhang2018q}
Z.~Zhang, L.~Ma, K.~K. Leung, L.~Tassiulas, and J.~Tucker, ``Q-placement:
  Reinforcement-learning-based service placement in software-defined
  networks,'' in \emph{2018 IEEE 38th International Conference on Distributed
  Computing Systems (ICDCS)}.

\bibitem{lin2016qos}
S.-C. Lin, I.~F. Akyildiz, P.~Wang, and M.~Luo, ``Qos-aware adaptive routing in
  multi-layer hierarchical software defined networks: a reinforcement learning
  approach,'' in \emph{2016 IEEE International Conference on Services Computing
  (SCC)}.

\end{thebibliography}

\vspace{12pt}

\end{document}